\documentclass[times,manuscript,twocolumn]{aastex701}
\usepackage[utf8]{inputenc}
\usepackage{xspace}

\usepackage{float}
\usepackage{graphicx}

\usepackage{amsmath}
\usepackage{xfrac}
\begin{document}

\title{Rate coefficients for dielectronic recombination of N-like Ne}

\author[0009-0002-5487-2042]{E.-O. Hanu}
\affiliation{GSI Helmholtzzentrum für Schwerionenforschung GmbH, 64291 Darmstadt, Germany}
\affiliation{Helmholtz-Institut Jena, 07743 Jena, Germany}
\affiliation{Institut für Kernphysik, Goethe Universität, 60438 Frankfurt am Main, Germany}
\email[show]{e.hanu@gsi.de}

\author[0000-0001-7384-1917]{M. Lestinsky}
\affiliation{GSI Helmholtzzentrum für Schwerionenforschung GmbH, 64291 Darmstadt, Germany}
\email{m.lestinsky@gsi.de}

\author[0000-0003-1451-7089]{E. B. Menz}
\affiliation{GSI Helmholtzzentrum für Schwerionenforschung GmbH, 64291 Darmstadt, Germany}
\affiliation{Institut für Kernphysik, Universität zu Köln, 50937 Köln, Germany}
\email{e.menz@gsi.de}

\author[0009-0009-7009-3859]{M. Looshorn}
\affiliation{I. Physikalisches Institut, Justus-Liebig-Universität, 35392 Gießen, Germany}
\affiliation{Helmholtz Forschungsakedemie Hessen für FAIR (HFHF), GSI Helmholtzzentrum\linebreak für Schwerionenforschung GmbH, 64291 Darmstadt, Germany}
\email{mirko-dieter.looshorn@physik.uni-giessen.de}

\author[0009-0003-0576-4317]{Z. Andelkovic}
\affiliation{GSI Helmholtzzentrum für Schwerionenforschung GmbH, 64291 Darmstadt, Germany}
\email{z.andelkovic@gsi.de}

\author[0009-0007-3777-1369]{A. Biniskos}
\affiliation{Institut für Kernphysik, Goethe Universität, 60438 Frankfurt am Main, Germany}
\email{a.biniskos@gsi.de}

\author[0000-0001-8825-8820]{C. Brandau}
\affiliation{GSI Helmholtzzentrum für Schwerionenforschung GmbH, 64291 Darmstadt, Germany}
\email{c.brandau@gsi.de}

\author{A. Bräuning-Demian}
\affiliation{GSI Helmholtzzentrum für Schwerionenforschung GmbH, 64291 Darmstadt, Germany}
\email{a.braeuning-demian@gsi.de}

\author[0000-0003-3511-262X,gname="M. R.",sname="Fogle"]{M. R. Fogle}
\affiliation{Physics Department, Auburn University, Auburn, AL 36849, USA}
\email{mrf0006@auburn.edu}

\author[0000-0003-1326-9394]{W. Geithner}
\affiliation{GSI Helmholtzzentrum für Schwerionenforschung GmbH, 64291 Darmstadt, Germany}
\email{w.geithner@gsi.de}

\author[0009-0001-4149-7100]{F. Herfurth}
\affiliation{GSI Helmholtzzentrum für Schwerionenforschung GmbH, 64291 Darmstadt, Germany}
\email{f.herfurth@gsi.de}

\author[0000-0003-0166-2666]{P.-M. Hillenbrand}
\affiliation{GSI Helmholtzzentrum für Schwerionenforschung GmbH, 64291 Darmstadt, Germany}
\email{p.m.hillenbrand@gsi.de}

\author[0000-0003-1727-8319]{C. Krantz}
\affiliation{GSI Helmholtzzentrum für Schwerionenforschung GmbH, 64291 Darmstadt, Germany}
\email{c.krantz@gsi.de}

\author[0000-0002-5843-0586]{R. Schuch}
\affiliation{Fysikum, Stockholm University, SE-106 91 Stockholm, Sweden}
\email{schuch@fysik.su.se}

\author[0009-0002-4338-8140]{M. Tatsch}
\affiliation{I. Physikalisches Institut, Justus-Liebig-Universität, 35392 Gießen, Germany}
\affiliation{Helmholtz Forschungsakedemie Hessen for FAIR (HFHF), GSI Helmholtzzentrum\linebreak für Schwerionenforschung GmbH, 64291 Darmstadt, Germany}
\email{maria.d.tatsch@lehramt.uni-giessen.de}

\author{G. Vorobyev}
\affiliation{GSI Helmholtzzentrum für Schwerionenforschung GmbH, 64291 Darmstadt, Germany}
\email{g.vorobyev@gsi.de}

\author[0000-0002-6305-3762]{S.-X. Wang}
\affiliation{I. Physikalisches Institut, Justus-Liebig-Universität, 35392 Gießen, Germany}
\affiliation{Helmholtz Forschungsakedemie Hessen for FAIR (HFHF), GSI Helmholtzzentrum\linebreak für Schwerionenforschung GmbH, 64291 Darmstadt, Germany}
\email{Shu-Xing.Wang@physik.uni-giessen.de}

\author[0000-0003-0461-3560]{T. St\"ohlker}
\affiliation{GSI Helmholtzzentrum für Schwerionenforschung GmbH, 64291 Darmstadt, Germany}
\affiliation{Helmholtz-Institut Jena, 07743 Jena, Germany}
\affiliation{Friedrich-Schiller-University Jena, 07743 Jena, Germany}
\email{t.stoehlker@gsi.de}

\author[0000-0002-6166-7138]{S. Schippers}
\affiliation{I. Physikalisches Institut, Justus-Liebig-Universität, 35392 Gießen, Germany}
\affiliation{Helmholtz Forschungsakedemie Hessen for FAIR (HFHF), GSI Helmholtzzentrum\linebreak für Schwerionenforschung GmbH, 64291 Darmstadt, Germany}
\email{schippers@jlug.de}

\begin{abstract}

Dielectronic recombination (DR) for the process Ne$^{3+}$ + e$^{-}$ $\rightarrow$ Ne$^{2+}$ was investigated in a merged-beams arrangement at the heavy-ion storage ring CRYRING@ESR. The energy-dependent DR rate coefficient, $\alpha(E)$ was measured over the electron-ion collision energy range from 0 to 25 eV. The measurements cover the complete set of DR resonance series associated with $2s\to2p$ core excitations. The primary ion beam is estimated to have consisted of $44\%$ of the ions in the ground state, with the remainder distributed among long-lived metastable levels. In addition to the measurements we carried out quantum mechanical calculations of DR cross sections. The theoretical treatment includes contributions from the ground and excited metastable initial levels, weighted according to the estimated beam composition. From the experimental energy-resolved spectra, we derive a temperature dependent DR plasma recombination rate coefficient $\alpha_\mathrm{exp}(T)$ (PRRC). In the temperature domain where Ne$^{3+}$ is abundant in collisionally ionized plasmas, the present results show a good agreement with the present and with previous theoretical predictions. In the low-temperature regime characteristic for photoionized plasmas, the experimentally derived DR plasma rate coefficient is slightly larger than the published theoretical ones and does not agree within the experimental uncertainties. Parametrized fits of the experimentally derived DR PRRC are presented in order to facilitate an easy inclusion into astrophysical modelling codes.

\end{abstract}


\section{Introduction} 

Planetary nebulae (PNe) represent one of the last stages in the lifecycle of aging stars with masses similar to the Sun.
Spectroscopic observations from space-based instruments enable quantitative measurements of PNe plasma emissions. This allows the identification of characteristic radiation from many elements in different ion charge states, abundant in the progenitor star's outer shell. Using known oscillator strengths, elemental abundances and charge state compositions can be derived from the observed spectra.

Systematic analyses of large PNe samples across the Milky Way revealed galactocentric radial gradients in elemental abundances, particularly for oxygen, neon, sulfur, and argon. These metallicity gradients reflect local stellar densities and nucleosynthetic histories and provide insight into galactic structure \citep{Bernard-Salas2008, Pagomenos2018}. 

Accurate interpretation of nebular emission spectra critically depends on reliable atomic data governing the ionization balance. In particular, this concerns recombination processes. Among these, dielectronic recombination (DR) plays a dominant role in setting the charge-state distribution of heavy elements in photoionized plasmas such as PNe. In a recent assessment of laboratory astrophysics priorities for NASA missions, \citet{Brickhouse2020} identified DR rate coefficients as a key data need for understanding nebular emission across a broad range of wavelengths, including the UV, optical, mid-infrared, EUV, and X-ray regimes. These DR data are essential for determining ionization parameters, ionization correction factors, and elemental abundances derived from spectroscopic observations. Despite their importance, large uncertainties persist in available DR data, particularly in the low-temperature regime most relevant to photoionized plasmas. Here, we address this gap by providing new DR measurements for Ne$^{3+}$, addressing long-standing DR data uncertainties that limit the reliability of ionization balance calculations in photoionized plasmas \citep{Savin2001}. The results offer experimental benchmarks for the theoretical models used in plasma simulations and strengthen the atomic foundation underlying Ne$^{3+}$ spectral diagnostics which is a step toward more reliable elemental abundances, metallicity gradients, and ionization conditions across astrophysical environments.

The low-temperature DR regime is especially problematic for photoionized environments such as PNe, X-ray binaries, and active galactic nuclei, where near-threshold resonances strongly influence recombination rates \citep{Brickhouse2020}. Small inaccuracies in theoretical resonance positions can translate into order-of-magnitude deviations in plasma rate coefficients \citep{Schippers2004}, directly impacting predicted charge-state distributions.

Neon is of special astrophysical importance due to its high cosmic abundance and its role as a tracer of chemical evolution and ionization conditions. It is the fifth most abundant element in the solar corona \citep{Asplund2009}, and its emission lines are commonly observed in PNe, H\,\textsc{ii} regions, and active galaxies \citep{Levesque2013, Meijerink2008, Zhuang2019}. Additionally, near-infrared spectroscopy of low-ionization neon has been proposed as a diagnostic tool for estimating star formation rates in active galaxies \citep{Zhuang2019}.

In particular, Ne$^{3+}$ is abundant in photoionized astrophysical plasmas, where its forbidden-line emission serves as a diagnostic of electron temperature and ionization conditions, and its high-ionization lines are a well-established tracer of AGN activity. The recent detection of the [Ne\,\textsc{iv}]\,$\lambda\lambda$2422,2424 $\mathring{A}$ forbidden doublet in the JWST/NIRSpec spectrum of GN-z11, an exceptionally luminous galaxy at $z = 10.6$, provided unambiguous evidence for an accreting supermassive black hole in the early universe \citep{Maiolino2024}. Because Ne$^{3+}$ emission requires ionizing photons with energies exceeding 63.5\,eV, it serves as a robust AGN diagnostic that is absent in star-forming galaxies. Accurate interpretation of such high-ionization nebular emission, including the derivation of ionization parameters and elemental abundances from observed spectra, critically depends on reliable DR rate coefficients for Ne$^{3+}$.

Several other neon charge states have been the subject of prior experimental DR studies. B-like Ne$^{5+}$ was studied at CRYRING in Stockholm \citep{Mahmood2013}. Be-like Ne$^{6+}$ was measured at the same storage ring \citep{Orban2008}. The Li-like Ne$^{7+}$ DR rate coefficient was also measured at CRYRING, covering $\Delta n = 0$ core excitations \citep{Zong1998, Boehm2005}. To date, the only published experimental DR measurement for a nitrogen-like ion is that of Fe$^{19+}$, covering $\Delta n = 0$ core excitations \citep{Savin2002}. To the best of our knowledge, no experimental DR measurement has been published for Ne$^{3+}$.

In this work, we report on a combined experimental and theoretical study in which we measured the absolute merged-beams recombination rate coefficient for the DR of Ne$^{3+}$, an ion with an N-like $2s^2 2p^3$ ground configuration. These measurements are compared with state-of-the-art theoretical predictions, allowing us to determine a plasma DR rate coefficient. Similar to previous publications by our group, we refer in the following to the Ne$^{3+}$ DR process in terms of the parent ion: the capture of an initially free electron into a continuum state by Ne$^{3+}$ leads to the formation of Ne$^{2+}$.

During this process, the captured Rydberg electron occupies an available $nl$ orbital while the excess energy from the capture is resonantly transferred onto one of the bound state electrons in the Ne$^{3+}$ core. The resulting [Ne$^{2+}$]** ion transiently forms a doubly excited, autoionizing state. This capture may proceed via multiple channels:

\begin{eqnarray}
\label{eqn:Ne3DRpathways}
\mathrm{Ne}^{3+}\ (2s^2\,2p^3\,) + e^{-}
  \leftrightarrow [\mathrm{Ne}^{2+}]^{**}\ \left\{\begin{array}{l}
    2s^2\,2p^3\ nl,\\
    2s\,2p^4\ nl.\\
  \end{array} \right.
\end{eqnarray}

Since the inner active electron remains in its original shell, DR via the indicated excited levels is referred to as $\Delta n = 0$ DR. With a certain probability, the doubly excited states formed in the dielectronic capture stabilize to below the autoionization threshold through the emission of photons, which can be observed as recombination lines in the Ne$^{2+}$ spectrum \citep{Cleri2023}.

\section{Experimental setup}

\begin{figure}[h!]
    \centering
    \includegraphics[width=0.44\textwidth]{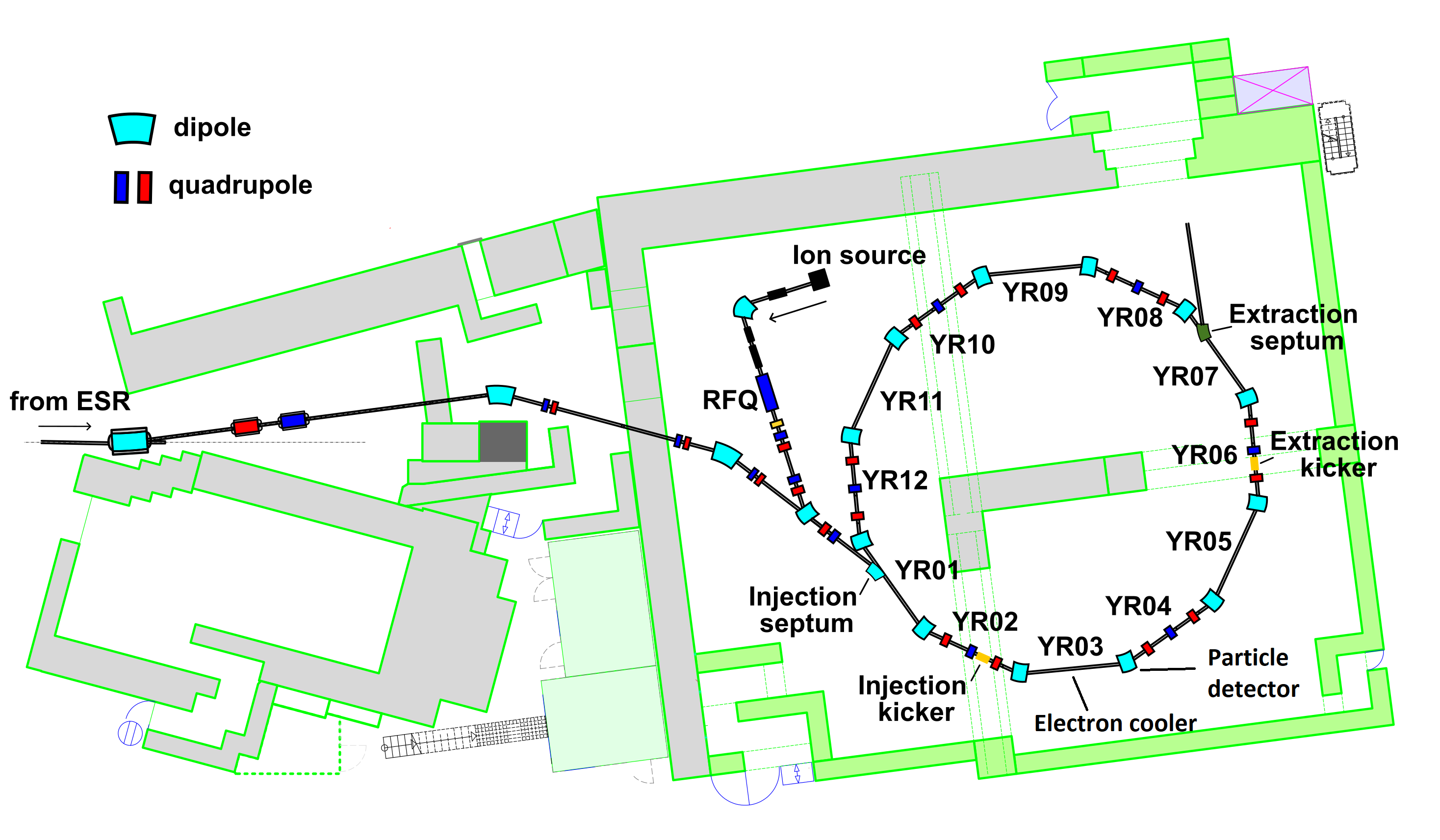}
    \caption{Schematic of CRYRING@ESR with its injection beamlines. The electron cooler and the detectors used for dielectronic recombination experiments are located in sections YR03 and YR04, respectively. For the present measurements, Ne$^{3+}$ were produced with the local ECR ion source \citep{Vorobyev2026}.}
    \label{fig:cryring_top_view}
\end{figure}

The low-energy heavy-ion storage ring CRYRING@ESR \citep{Lestinsky2016} has been integrated into the GSI/FAIR accelerator chain and is operating downstream of the ESR. A schematic of the ring is shown in \autoref{fig:cryring_top_view}.

CRYRING@ESR is equipped with an ultracold electron cooler, implementing adiabatic transverse expansion \citep{Danared2000} and has recently demonstrated its resolving power in dielectronic recombination experiments at the physical limits of the present design \citep{Krantz2026}. Recombination experiments at CRYRING@ESR have been described in earlier studies \citep{Brandau2025, Schippers2025, Krantz2026} and we just summarize the experimental parameters specific to the present Ne$^{3+}$ measurement. The ion beam was produced by an electron cyclotron-resonance (ECR) source \citep{Broetz2001, Vorobyev2026} and injected in the ring at an energy of 5.35 keV/u. The ions were then accelerated to a final energy of 2.23 MeV/u, and the stored beam's intensity was approximately 5.5 $\cdot$ 10$^6$ ions immediately after injection of an ion bunch into CRYRING@ESR.

\begin{figure*}[ht!]
    \centering
    \includegraphics[width=\textwidth]{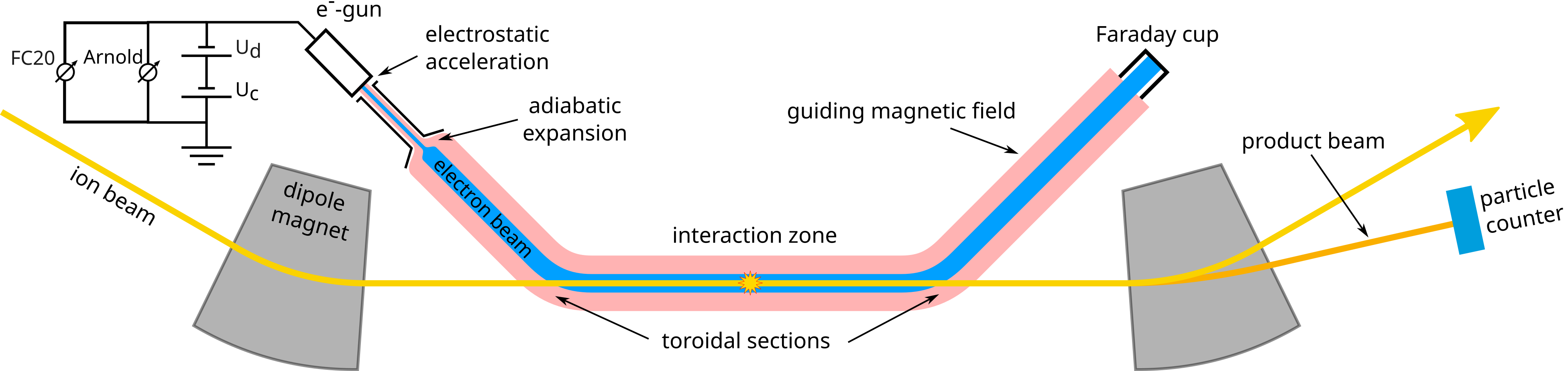}
    \caption{Sketch of the present electron-ion recombination experiment at the CRYRING@ESR electron cooler.}
    \label{fig:ecoolersetup}
\end{figure*}

Inside the electron cooler, the electron beam overlaps collinearly with the ion beam along the interaction region for a length of approximately 0.9 m (see \autoref{fig:ecoolersetup}). For the present experiment, the electron beam was adiabatically expanded with an expansion factor $\xi = 33$. The electron cooler serves a dual purpose: initially cooling the ion beam by reducing its transverse and longitudinal momentum spreads through interactions with the significantly colder electrons \citep{Poth1990}, and subsequently by acting as a target for electron-ion collisions, in which ions capture free electrons and change their charge state. The electron-ion collision energy was scanned across the theoretically predicted range by varying the electron energy accordingly.

The recombined Ne$^{2+}$ ions were separated from the beam by the first 30$^{\circ}$ dipole magnet downstream from the interaction region and directed into a movable particle detector. The detector employed a fast scintillation counter \citep{Hahn2020} capable of registering individual ions with near 100\% efficiency. It was positioned such that ions impacted near the center of its 20 mm wide active area. Scintillation light signals were recorded by a photomultiplier, amplified, pulse-shaped with a fast timing amplifier and converted into digital pulses with a leading-edge discriminator. The resulting digital pulses were counted by a scaler channel of the data acquisition system. The analog photomultiplier output was continuously monitored during the beamtime, and no degradation in detector performance was observed throughout the course of our experiment. The average count rates amounted to approximately 500 $\mathrm{s}^{-1}$ at the start of each cycle. The pulse lengths were of the order of 100 ns. Under these conditions, dead-time losses were negligible. So-called ``blind`` measurements, without ions in the ring, were performed periodically to determine the dark count rates of the detector systems connected to the data acquisition. For this purpose, the ion beam path was temporarily blocked by inserting a Faraday cup into the closed orbit of CRYRING@ESR, and data were recorded into separate files but otherwise under the same settings as the corresponding energy scan. Typically, blind measurements had a duration of five to ten minutes. The determined systematic offsets in detector count rate and ion beam current measurements were subsequently subtracted from the data. A dark count rate of 30 $\mathrm{s}^{-1}$ was measured for the particle detector and corrected for accordingly. Dark counts originated from ambient light sources, including the thermionic electron gun of the upstream electron cooler, nearby vacuum gauges, and emissions from ionization pumps.

The relative electron–ion collision energy was precisely controlled by varying the voltage of the cathode of the electron cooler \citep{Brandau2025}. Rapid voltage detunings were achieved using a bipolar high-voltage amplifier (KEPCO BOP 1000M), which generated voltage offsets $U_d$ on millisecond time scales and which was connected in series with the cooler cathode supply generating the voltage $U_c$  (see \autoref{fig:ecoolersetup}). This setup allows incremental scans across the energy range of interest, with a voltage range of $\pm$1000 V relative to the cooling voltage (at which the relative energy between the ion and electron beam is zero), enabling symmetric coverage around the zero relative-energy point.

Precise determination of the electron beam energy relied on high-voltage monitoring which was facilitated by two high-voltage probes named ``Arnold``  and ``FC20``, which were operated in parallel (see \autoref{fig:ecoolersetup}). The difference between the two devices is their accuracy. The newer FC20 divider \citep{Dirkes2020, Ueberholz2025} is expected to be an order of magnitude more accurate than the much older Arnold divider \citep{Brandau2025}, which currently mainly serves as a backup system.

During DR measurements, the electron energy was varied in an incremental sequence. At the beginning of each step, the voltage was held constant for a short period to allow the system to reach quasi-steady conditions before data acquisition commenced. Furthermore, intermediate cooling intervals were periodically interleaved to maintain the ion beam energy close to the nominal cooling condition. This measurement scheme enables accurate determination of recombination spectra while minimizing systematic effects associated with transient voltage changes \citep{Menz2024}.

In practice, the electron energy is scanned using the established 'wobble' mode measurement scheme \citep{Lestinsky2022, Brandau2025}. In this mode, the electron energy repeatedly cycles through a sequence of 'cooling', 'reference' and 'measurement' steps, each with a duration of 10 ms. In the following, this sequence is denoted as `c-r-m'. The cooling step restores the nominal cooling condition to preserve the ion-beam quality during the scan and the cathode potential is adjusted such that electrons and ions have the same mean velocity. The reference step is performed at a fixed relative energy, ideally within the overall scan range but chosen such that the merged-beams rate coefficient at the reference energy is free from dielectronic resonances and reflects only a small contribution from nonresonant radiative recombination (RR). Under these conditions, the detected count rate is dominated by background events, primarily arising from charge-transfer collisions of the stored ions with residual gas particles. The count rate measured during the reference step therefore provides a background signal that can be subtracted from the measurement step count rate. The small RR contribution contained in this background can be estimated with sufficient accuracy from theory and subsequently added back to the corrected measurement signal \citep{Schippers2001}. During each measurement step, an incrementally adjusted voltage detuning $U_d$ is applied to the acceleration voltage $U_c + U_d$ via the KEPCO HV amplifier for typically $\sim$10\,ms, during which the particle detector records the number of ions undergoing electron capture. Absolute recombination rate coefficients were derived from the measured count rates as a function of electron–ion collision energy as detailed by \citet{Brandau2025}.

\section{Derivation of merged beam recombination rate coefficient}

\begin{figure*}[ht]
    \hspace*{-1.4cm}
    \includegraphics[width=1.15\textwidth]{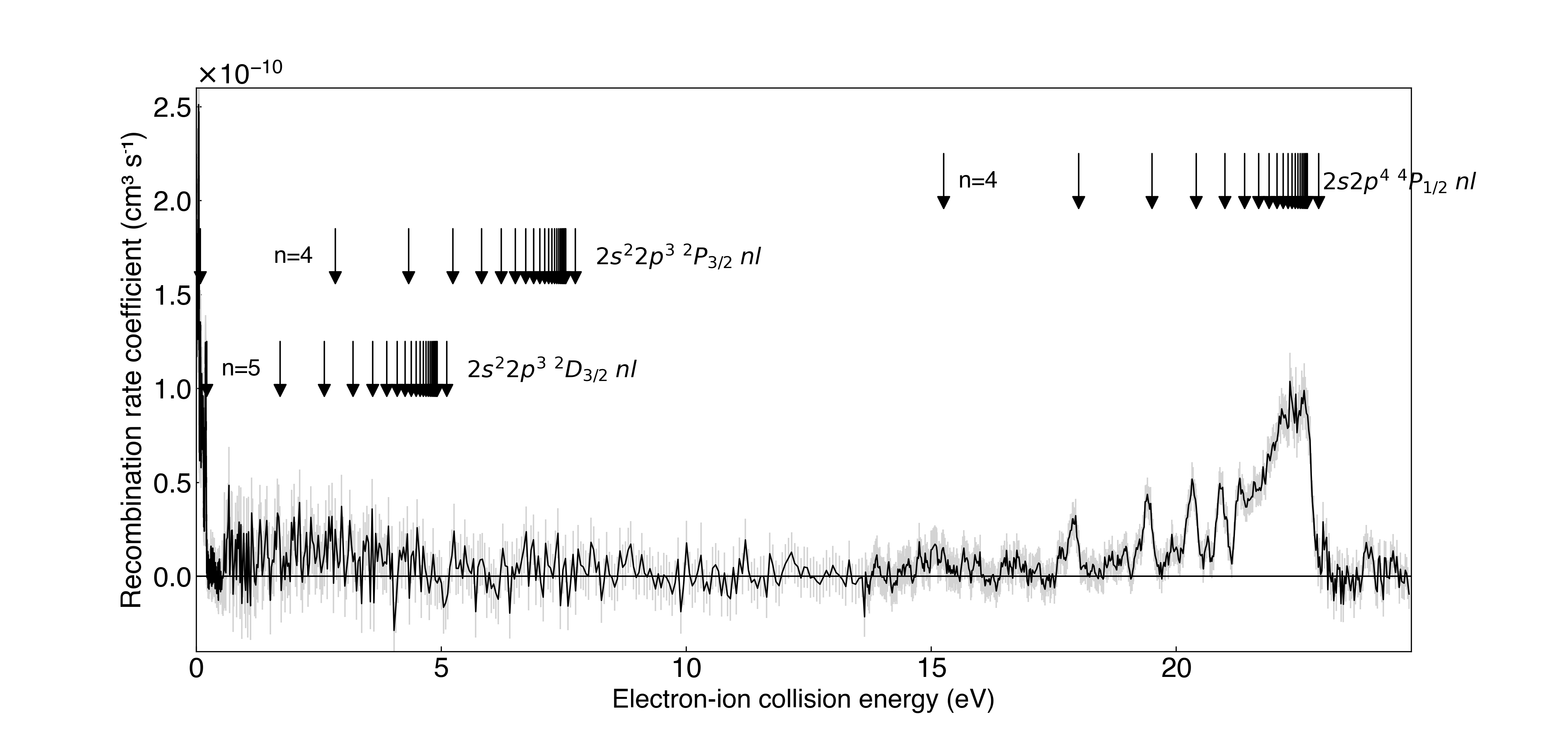}
    \caption{Overview of the measured and deconvolved MBRRC $\alpha(E_\mathrm{cm})$ for DR of Ne$^{3+}$, covering a collision energy range between 0 and 25 eV. The experimental data are depicted as a black line, with the grey shading representing the associated statistical and systematic uncertainties. To facilitate interpretation, the resonance positions of three notable DR series are marked by downward-pointing arrows, derived from the Rydberg formula (see \autoref{eqn:rydberg_formula}), while the last arrows denote the corresponding $n \rightarrow \infty$ series convergence limits.}
    \label{fig:overview_spectrum}
\end{figure*}

The experimental merged-beams recombination rate coefficient (MBRRC) for radiative recombination (RR) and dielectronic recombination (DR) of Ne$^{3+}$ is shown in \autoref{fig:overview_spectrum}, which covers a center-of-mass energy range of 0 - 25 eV. 

The observed $\Delta n$ = 0 DR resonances in this range correspond to core excitations from the $1s^2\,2s^2\,2p^3$ $^4S$ ground state to the $1s^2\,2s^2\,2p^3$ $^2D_{3/2}$, $^2P_{3/2}$ and $1s^2\,2s\,2p^4$ $^4P_{1/2}$ excited states with the capture of the initially free electron into a high-$n$  Rydberg level, with $n\rightarrow \infty$ series limits at $E_{\infty}=$ 5.11, 7.74, and 22.91 eV \citep{ASD}. The positions of the resonances belonging to each Rydberg series are indicated by vertical arrows in \autoref{fig:overview_spectrum}. They can be approximated by the Bohr formula for an ion of charge $q$:

\begin{equation}
    E_n \approx E_{\infty} - 13.606\ \text{eV} \cdot \frac{q^2}{n^2},
    \label{eqn:rydberg_formula}
\end{equation}

with $q=3$ for Ne$^{3+}$. This simplified hydrogenic approximation neglects any core-electron interactions.

The merged-beams recombination rate coefficient $\alpha(E_\mathrm{cm})$ as a function of the electron-ion collision energy $E_\mathrm{cm}$ in the electron-ion center of mass frame is normalized to an absolute scale by dividing the background-subtracted count rate $R(E_\mathrm{cm})$ by the stored ion number $N_i$ and electron density $n_e$, as given by the following formula \citep{Wang2024}:

\begin{equation}
    \alpha_\mathrm{exp}(E_\mathrm{cm}) = \frac{R(E_\mathrm{cm}) C}{(1-\beta_e(E_\mathrm{cm}) \beta_i) N_i n_e(E_\mathrm{cm}) \eta L}
\end{equation}

Here, $C$ = 54.178 m denotes the storage ring circumference, $L$ = 0.9 m the length of the straight section of the electron beam \citep{Brandau2025} and $\eta$ the single-particle detector efficiency. In this experiment, $\beta_i$ is held constant while $\beta_e$ is tuned by adjusting $E_\mathrm{cm}$ through the cathode voltage of the electron gun. The number $N_i$ of stored ions follows from the ion current $I_\mathrm{ion}$ according to:

\begin{equation}
    N_i = \frac{I_\mathrm{ion}}{e q f_\mathrm{rev}}
\end{equation}

with $f_\mathrm{rev}$ being the revolution frequency. Finally, the electron density is obtained from the electron current $I_e$ via:
\begin{equation}
    n_e = \frac{I_e}{e \xi \pi r^2_\mathrm{cath} \beta_e c}
\end{equation}
where $\xi = 33$ is the electron beam expansion factor and $r_\mathrm{cath}$ = 2 mm is the cathode radius. Since $\beta_e$ and the electron density $n_e$ are correlated parameters, connected through the space charge of the electron beam, an iterative procedure is applied to derive these quantities \citep{Schippers2001}.

Due to the relatively short ion beam lifetime of 19 s, we divided the complete energy range of interest into several scan intervals each covered by separate runs. The shown composite spectrum represents the sum of the measurements from the separate intervals (see \autoref{fig:overview_spectrum}).

The merged-beams configuration used at the electron cooler introduces a systematic distortion of the measured DR spectrum, arising from the toroidal magnetic fields used to merge and de-merge the electron beam from the ion beam. In the overlap regions where the electron beam and the ion beam are not collinear, the set acceleration voltage corresponds to a higher relative energy than in the straight section. This effect is corrected by deconvolving the cooler electron energy distribution from the measured recombination rate coefficient using the hydrocal code \citep{hydrocal}, which implements the deconvolution procedure described by \citet{Lampert1996}.

Our DR spectrum for Ne$^{3+}$ is given in \autoref{fig:overview_spectrum} with statistical errors and after RR (radiative recombination) substraction and deconvolution. Systematic uncertainties arise mainly from a limited knowledge of the electron beam profile and the ion current measurement. The overlap geometry of the electron-ion beam interaction region, as given in \citet{Brandau2025}. The total systematic uncertainty of the absolute recombination rate coefficient, obtained by adding all individual contributions in quadrature, amounts to $\approx \pm 15\%$ at the $1\sigma$ confidence level \citep{Lestinsky2022}.

\subsection{Estimating the metastable composition in the Ne$^{3+}$ parent ion beam}

The Ne$^{3+}$ ion possesses two long-lived metastable levels, $^2D_{5/2}$ ($\tau \approx 1719$ s) and $^2D_{3/2}$ ($\tau \approx 182$ s) \citep{Froese-Fischer_NIST_database}, with lifetimes far exceeding the stored ion beam lifetime of 19 seconds. Consequently, a substantial fraction of the Ne$^{3+}$ ions in the beam is expected to be in metastable excited levels throughout the experiment. Since the recombination cross sections depend sensitively on the initial quantum state of the recombining ion, a reliable estimate of the metastable population fractions in the stored beam is essential for a correct interpretation of the measured data.

To this end, a model was employed to describe the metastable beam composition of Ne$^{3+}$ ions formed by hot electron impact ionization in an ECRIS and to follow its temporal evolution after their extraction from the source. The initial level populations are chosen by assuming that metastable states are populated via electron-ion collisions in the ion source, following a thermal distribution weighted by the statistical degeneracy of each atomic level. Upon extraction from the electron cyclotron resonance (ECR) ion source at the time $t=0$, the subsequent evolution of the level populations is governed solely by a chain of radiative decays, and is modelled using a rate equation network for spontaneous emission from excited levels \citep{Lestinsky2012b} and using the A$_{kj}$ Einstein rate coefficients for spontaneous emission from \citet{Froese-Fischer_NIST_database}.

\begin{figure}[h]
    \hspace*{-0.9cm}
    \includegraphics[width=1.17\columnwidth]{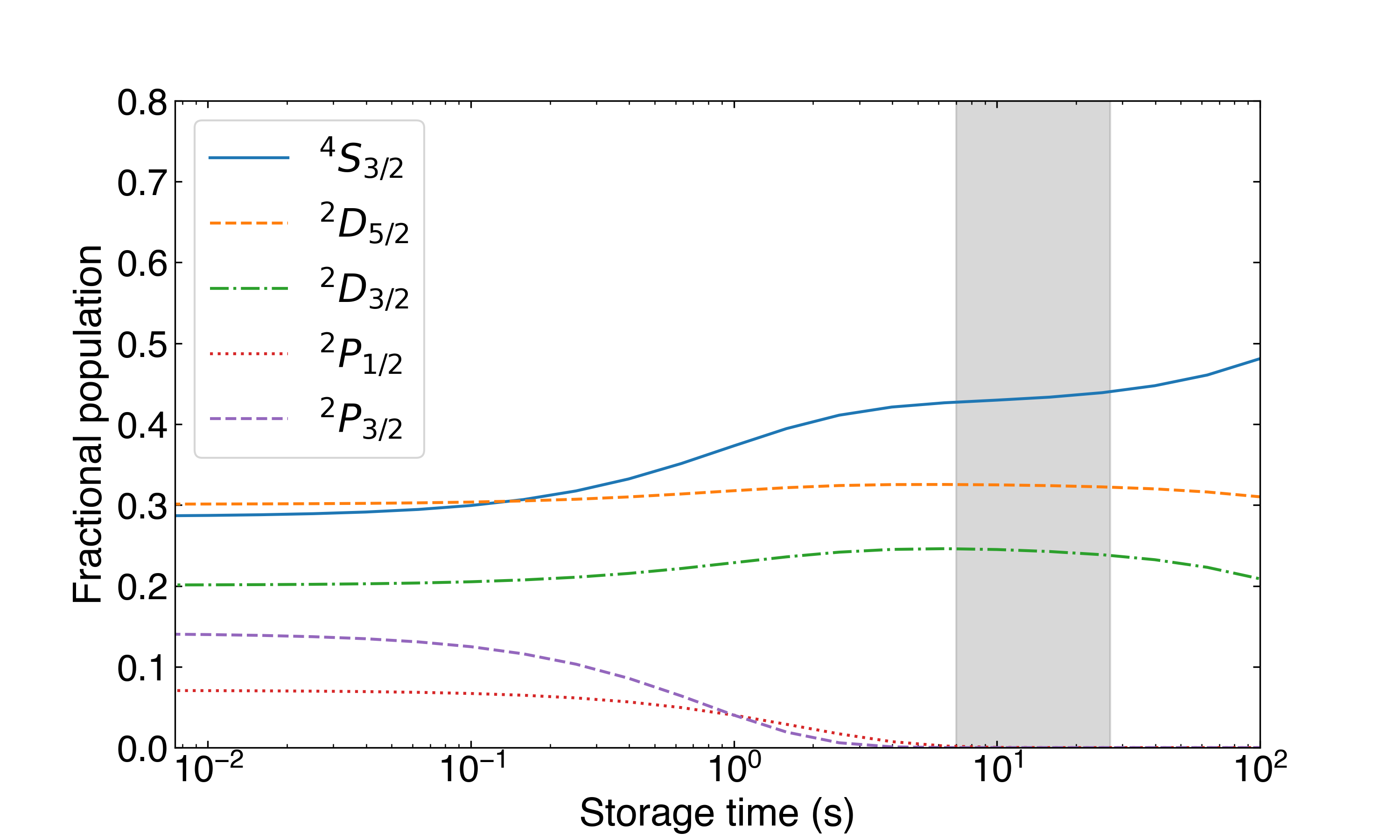}
    \caption{Evolution of Ne$^{3+}$ beam fractions in excited and metastable levels as a function of time. The grey shaded area marks the time range of the DR scan (7 to 27 s).}
    \label{fig:fractionalpopulation}
\end{figure}

The results of the model calculation are presented in \autoref{fig:fractionalpopulation}. It is noted that the ground state becomes the dominant population component after only 0.1 s. Considering the times required for injection, acceleration and initial electron cooling, the DR scan was performed on stored ion ensembles whose composition evolved for 7 to 27 s after extraction. For this time range we find that approximately 44\% of the total ions in the beam are in the ground level ($^4S_{3/2}$). The low-lying metastable levels account for the remainder of the ion population, with $\approx 32\%$ in $^2D_{5/2}$ and $\approx24\%$ in $^2D_{3/2}$.

\begin{figure}[h!]
    \hspace*{-0.8cm}
    \includegraphics[width=1.15\columnwidth]{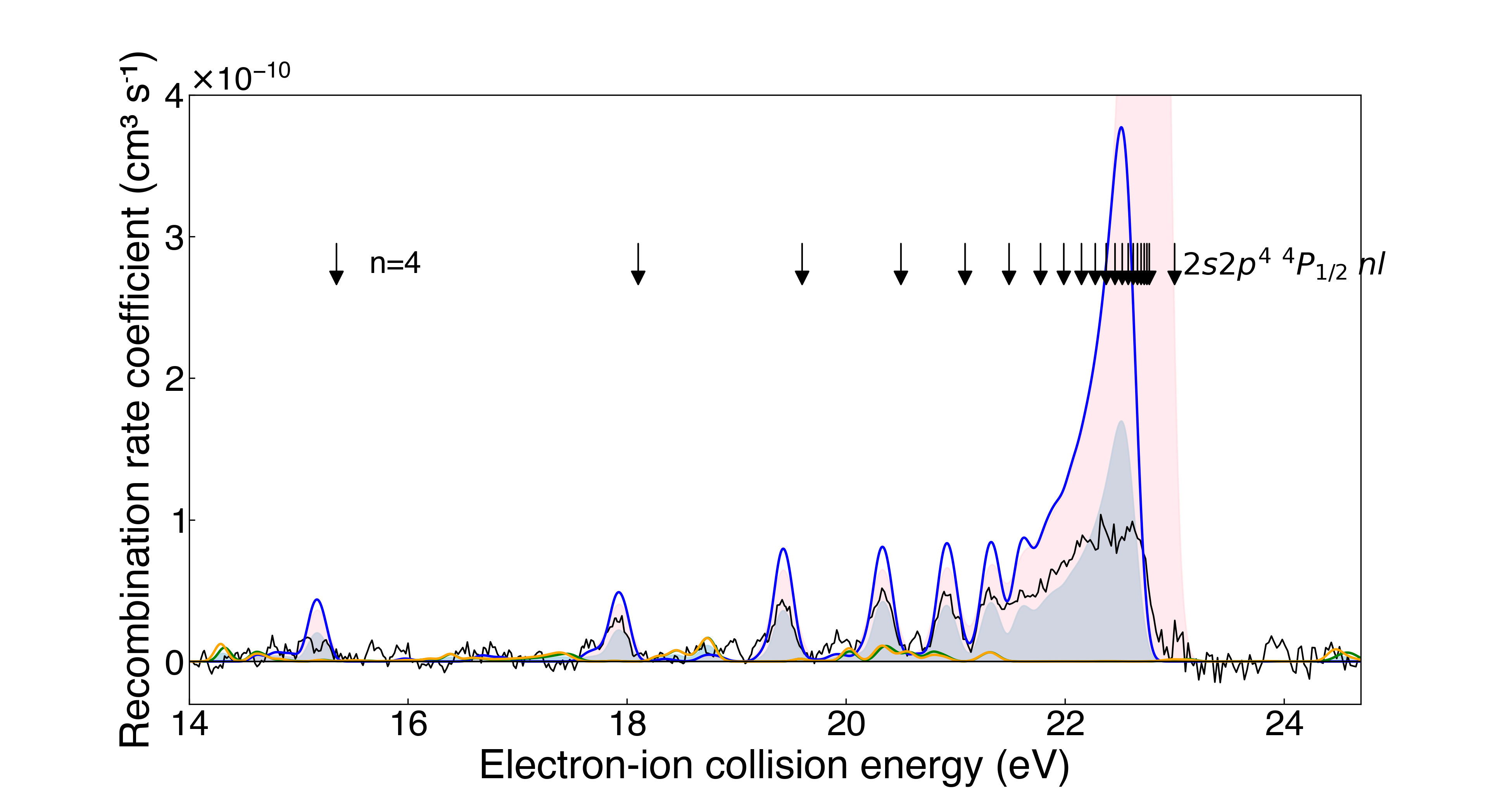}
    \caption{\label{fig:ne3mbrrctheory}The merged-beams DR rate coefficient $\alpha(E_{cm})$ shown for collision energies between 14 and 25 eV. The deconvolved and RR subtracted experimental result is depicted as the black line. Individual \textsc{autostructure} calculations for each metastable state are presented (in blue: $^4S_{3/2}$, in orange: $^2D_{5/2}$, in green: $^2D_{3/2}$). The weighted sum of all the states comprising the beam is plotted as a grey shaded curve. The pink shaded curve represents the theoretical result including all DR resonance strength up to $n$ = 1000.}
\end{figure}

In \autoref{fig:ne3mbrrctheory}, the experimental MBRRC is compared with results from quantum mechanical DR calculations using the \textsc{autostructure} code \citep{Badnell2011} for the ground and metastable levels for the collision energy region between 14 and 24.5 eV encompassing the field ionization region of the spectrum. The individual contributions of the $^4S_{3/2}$ (blue curve), $^2D_{5/2}$ (orange curve) and $^2D_{3/2}$ (green curve) levels have been weighted according to the ion beam composition from the model calculation (see \autoref{fig:fractionalpopulation}). The weighted sum is shown as the grey-shaded area. The pink shaded area represents the theoretical DR rate coefficient including the full resonance strength up to $n=1000$, thus accounting for Rydberg resonance contributions beyond the experimental field-ionization cut-off at $n_\mathrm{cut}=23$. This cut-off arises because Rydberg states with $n \gtrsim 23$ are too weakly bound to survive the motional electric field in the first downstream dipole magnet.

In the energy range between 1 and 9 eV (see \autoref{fig:overview_spectrum}), the experimental MBRRC exhibits multiple blended resonance features for which a rigorous quantitative analysis remains challenging given the current level of statistical uncertainty. Nevertheless, a reasonably good overall agreement is observed between the measured and theoretically calculated resonance energy structures in the region above 9 eV for the strengths of $\Delta n = 0$ DR.

Significant differences remain in several important spectral regions. These differences are derived by integration of both data sets in energy ranges where the respective spectrum cut feature distinct resonances up to the hard cut-off $n_\mathrm{cut}$ for field ionization. Particularly the strength of the rate coefficient associated with the energy range between 0.5 and 17.0 eV is considerably underpredicted by theoretical calculations even considering the combined statistical and systematic uncertainties of the experiment.
For each chosen energy range, the integrated theoretical and experimental rate coefficients and their associated statistical uncertainty are given in \autoref{table:strengthdifferences}.

\begin{table}[h]
\caption{Integrated DR rate coefficients for Ne$^{3+}$ ion. Differences in integrated DR resonance strengths between theory and experiment for important parts of the MBRRC spectrum, including statistical uncertainties for each interval. Here, $\kappa = \sfrac{ \int \alpha_\mathrm{theo} dE}{\int \alpha_\mathrm{exp} dE}$.}
\label{table:strengthdifferences}
\centering
\footnotesize
\hspace*{-1.5cm}
\begin{tabular}{llll}
\tableline
\multicolumn{1}{c}{Energy range} & $\int \alpha_\mathrm{theo} dE$ & $\int \alpha_\mathrm{exp} dE$ & $\kappa$ (stat. unc.) \\
\multicolumn{1}{c}{(eV)} & ($\mathrm{cm^3\,s^{-1}\,eV}$) & ($\mathrm{cm^3\,s^{-1}\,eV}$) & \multicolumn{1}{c}{(percent)}\\
\tableline
0 - 0.5   & $6.4913 \cdot 10^{-12}$ & $2.362 \cdot 10^{-11}$ & $0.274\ (0.84\%)$ \\
0.5 - 17.0  & $2.978 \cdot 10^{-11}$ & $7.324 \cdot 10^{-11}$ & $0.406\ (14.43\%)$ \\
17.0 - 22.0 & $6.821 \cdot 10^{-11}$ & $9.644 \cdot 10^{-11}$ & $0.707\ (2.72\%)$ \\
22.0 - 23.5 & $7.514 \cdot 10^{-11}$ & $6.857 \cdot 10^{-11}$ & $1.095\ (3.11\%)$ \\
\tableline
\tableline
\end{tabular}
\end{table}

\subsection{Recombination at low energies}

\begin{figure}[h]
    \hspace*{-0.8cm}
    \includegraphics[width=1.19\columnwidth]{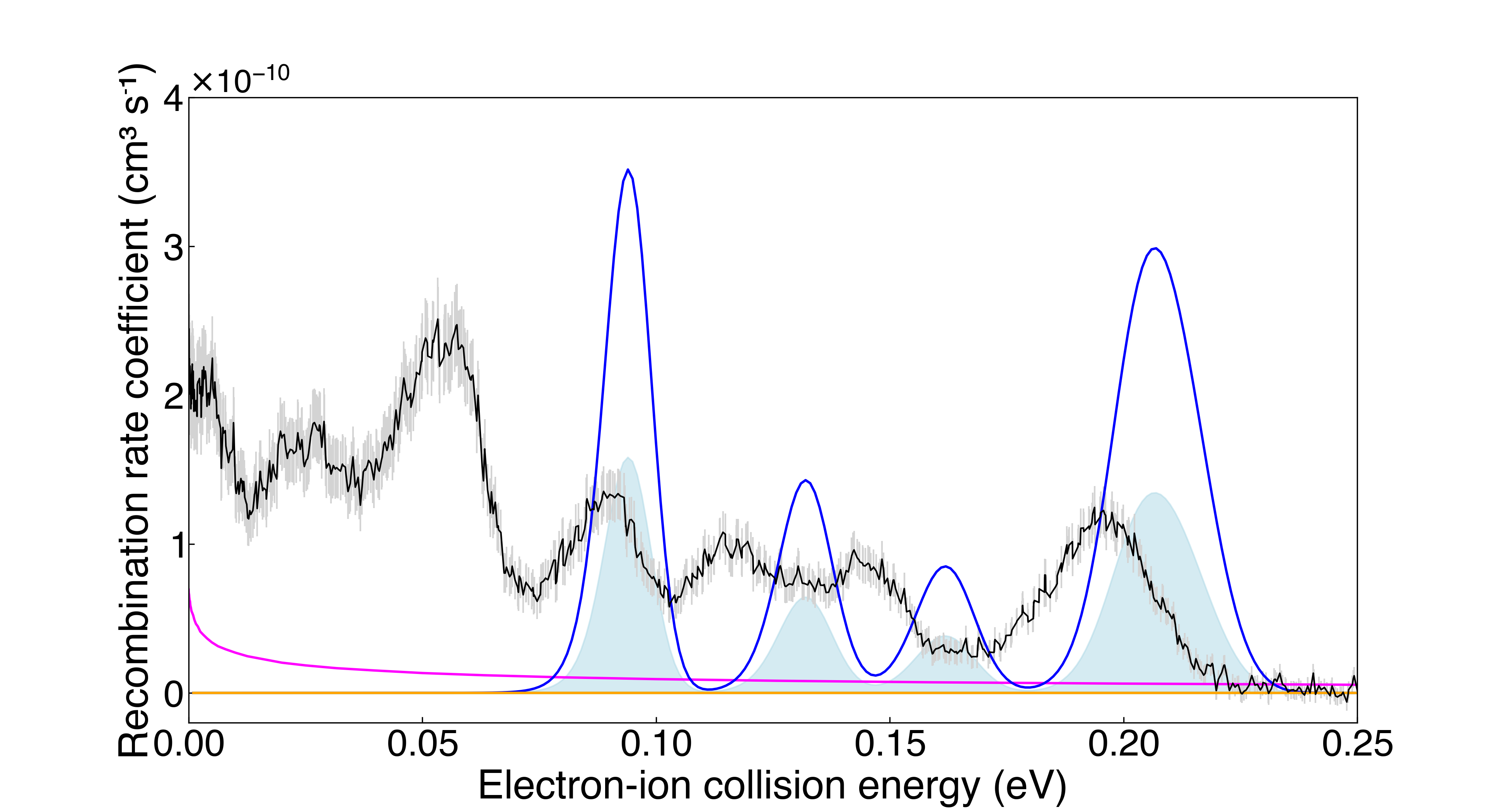}
        \caption{Deconvolved experimental data in the energy range of the DR resonances close to zero electron-ion collision energy. Line styles and colors follow the same convention as in \autoref{fig:ne3mbrrctheory}. Additionally, the magenta line represents the calculated RR rate coefficient.}
    \label{fig:ne3+lowlog}
\end{figure}

The experimental MBRRC covering the energy range from 0 to 0.25 eV is shown in \autoref{fig:ne3+lowlog}. Throughout this interval, the recombination rate coefficient is predominantly governed by DR resonances arising from core excitations of the $1s^2\, ($2s$^2\,$2p$^3$ $^4S$) ground state to the $1s^2\, ($2s$\,$2p$^4$ $^4P$) excited states, with capture into high-$\mathrm{n}$ Rydberg levels. We observe a significant discrepancy of the \textsc{autostructure} calculations with our experimental results.

\begin{figure}[h]
    \hspace*{-0.5cm}
    \includegraphics[width=1.05\columnwidth]{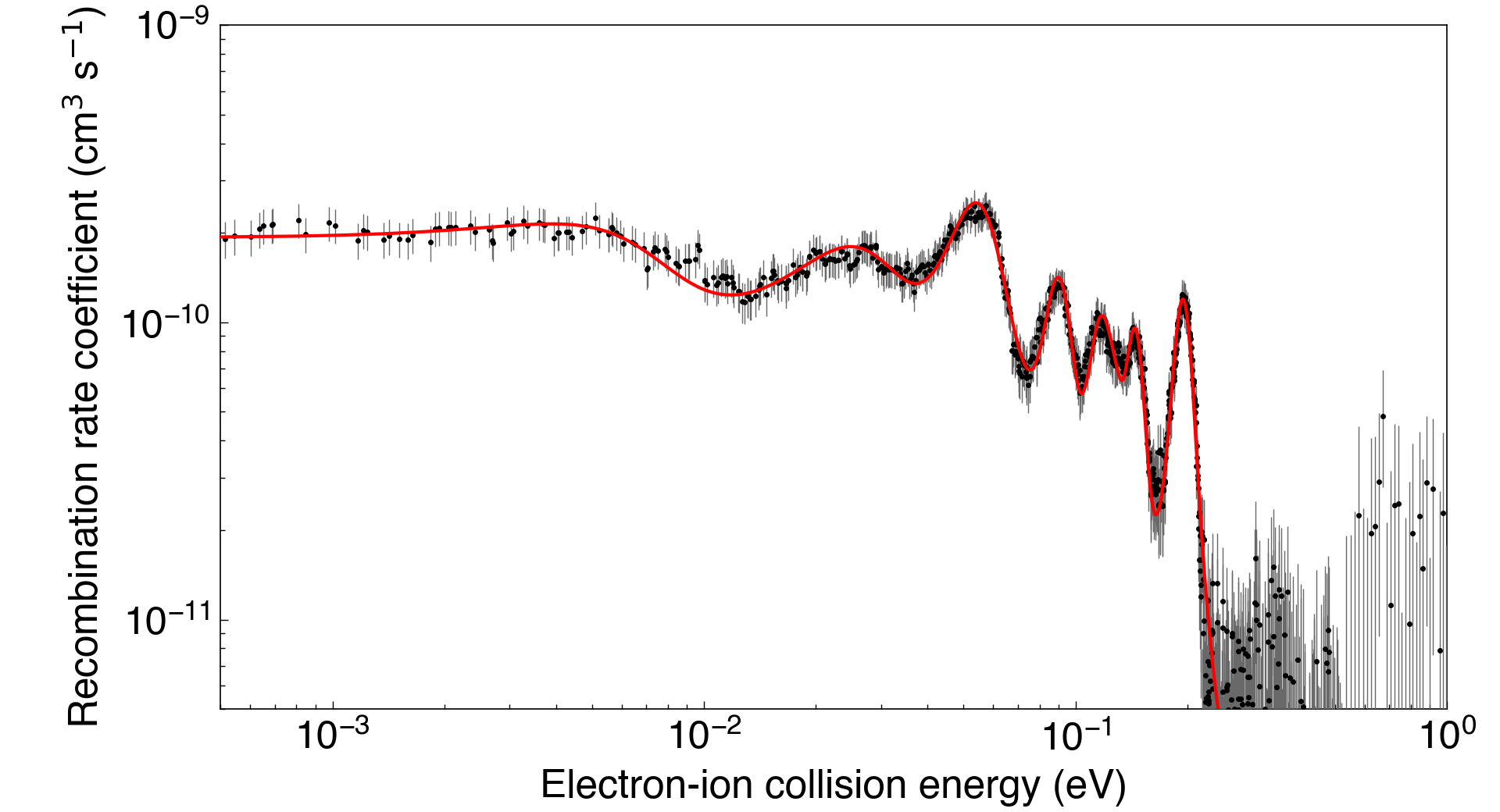}
    \caption{MBRRC in the energy range of the DR resonances close to zero electron-ion collision energy: experiment with the associated uncertainty (black bars) and empirical fit of model DR resonances (red line) as given in \autoref{table:lowresfit}.}
    \label{fig:ne3+fit}
\end{figure}

To accurately determine the resonant part of the recombination cross sections, eight DR resonances were fitted to the measured spectrum, and the corresponding cross sections were convolved with the experimental electron energy distribution. For the 0 -- 0.25 eV energy region of the spectrum, a synthetic model cross section $\hat{\sigma}(E_\mathrm{cm})$ was constructed, encompassing contributions from radiative recombination and eight partially overlapping DR resonances, some exhibiting substantial Lorentzian linewidths. Convolution of this model cross section with the flattened Maxwellian electron energy distribution characteristic of the experiment yields the corresponding fitted MBRRC: $\hat{\alpha}(E_\mathrm{cm}) = \left< v\hat{\sigma}(E_\mathrm{cm}) \right>$ where $v$ is the relative electron-ion velocity. The experimental $\alpha_\mathrm{exp} (E_\mathrm{cm})$ alongside the fitted synthetic MBRRC $\hat{\alpha}(E_\mathrm{cm})$ are displayed over the energy region from 0.5 meV to 1 eV in \autoref{fig:ne3+fit}.  The experimental points below 1 meV were excluded from the fit. The general fitting procedure has been described previously \citep{Schippers2004}.

The resonance parameters extracted from the fitting procedure are summarized in \autoref{table:lowresfit}.

\begin{table}[h!]
\caption{Parameters of the model cross section fitted to the DR spectrum below 0.25 eV, with resonance energy (eV), strength and Lorentzian width (eV).}
\label{table:lowresfit}
\hspace*{-1cm}
\begin{tabular}{llll}
\tableline
\multicolumn{1}{c}{Resonance} & Energy & Strength & Width\\
  & (eV) & (cm$^2$ eV) & (eV)\\ 
\tableline
1 & 0.0076 & $3.295\cdot 10^{-19}$  & $5.5\cdot 10^{-3}$ \\
2 & 0.029 & $5.0607\cdot 10^{-19}$  & $2\cdot 10^{-2}$ \\
3 & 0.058 & $5.2314\cdot 10^{-19}$  &  $1.7\cdot 10^{-2}$ \\
4 & 0.093 & $1.4252\cdot 10^{-19}$  & $8\cdot 10^{-3}$ \\
5 & 0.12 & $1.0974\cdot 10^{-19}$ & $10^{-2}$ \\
6 & 0.13 & $1.133\cdot 10^{-20}$ & $10^{-4}$ \\
7 & 0.147 & $8.5504\cdot 10^{-20}$ & $7\cdot 10^{-3}$ \\
8 & 0.197 & $1.5535\cdot 10^{-19}$ & $1.5\cdot 10^{-2}$ \\
\tableline
\tableline
\end{tabular}
\end{table}

\section{Plasma recombination rate coefficients}

\subsection{Derivation}

The DR plasma recombination rate coefficient (PRRC) is the temperature-dependent convolution of the DR cross section with a Maxwellian electron energy distribution.  It expresses the recombination rate per unit volume normalized to the electron and ion number densities for a plasma at a given electron temperature.

To obtain the plasma rate coefficient from the experimental MBRRC, several aspects related to the measured data must be considered. To this end, we followed the procedure outlined by \citet{Schmidt2006} and \citet{Lestinsky2009} for deriving the experimentally determined DR plasma recombination rate coefficient, $\alpha_{\mathrm{exp}}(T)$, from our measured data $\alpha_{\mathrm{exp}}(E_\mathrm{cm})$. The plasma recombination rate coefficient is formally obtained by convolving the energy-dependent recombination cross section with an isotropic Maxwellian electron velocity distribution function. This yields:

\begin{equation}
    \alpha_\mathrm{exp}(T) = \frac{4}{\sqrt{2\pi m_e}(k T)^{3/2}} \int_{0}^{\infty} \sigma(E) E \exp \left(-\frac{E}{k T}\right) dE
\label{eqn:convolution}
\end{equation}

The temperature regimes in which Ne$^{3+}$ is present in a plasma depend on the dominant formation mechanisms and the plasma temperature. Two idealized plasma models are commonly considered: a collisionally ionized plasma (CP) and an optically thin, photoionized plasma (PP). The relevant temperature ranges, defined as the temperature intervals over which Ne$^{3+}$ exceeds a fractional abundance of 1\% ($[\text{Ne}^{3+}]/[\text{Ne}] \ge 0.01$) in the charge state balance, were extracted graphically from \citet{Kallman&Bautista2001} for PP ($0.68 \le kT \ge 1.46$ eV) and from \citet{Bryans2009} for CP ($6.8 \le kT \ge 34$ eV). We indicate both corresponding temperature regimes in the plasma recombination rate coefficient figures as grey shaded regions in the following figures.

\subsection{Results and Comparison}

\begin{figure*}[ht]
     \hspace*{-1.2cm}
    \includegraphics[width=1.16\textwidth]{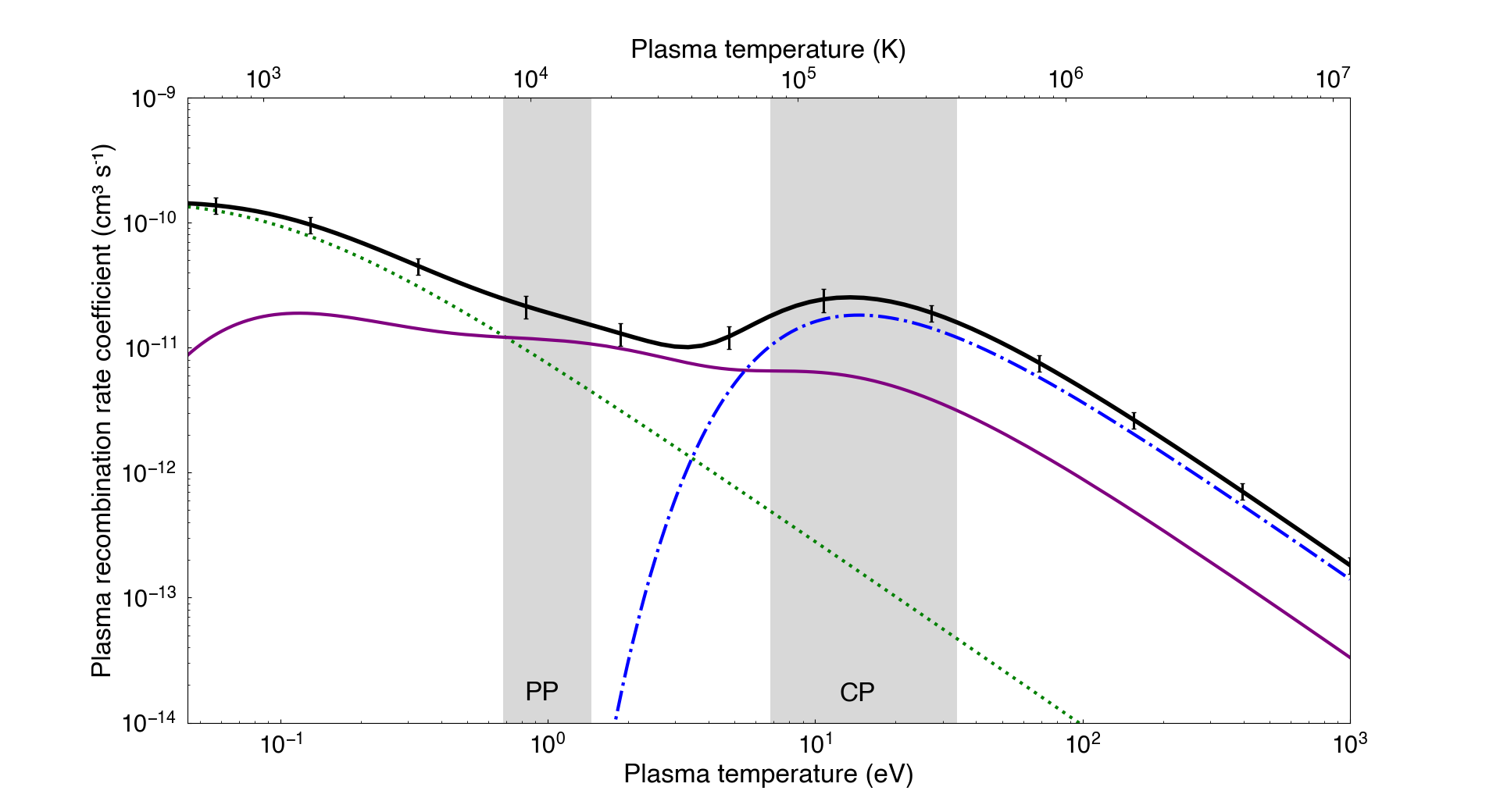}
    \caption{The experimentally derived total DR PRRC for Ne$^{3+}$ is represented by the black line. The error bars represent the total uncertainty (quadrature sum of statistical and systematic uncertainties). The individual components constituting the experimentally derived DR PRRC are also shown separately: the low-energy resonance fit (dashed green line), the contribution of the experimental spectrum (\autoref{fig:overview_spectrum}) in the energy range 0.25 -- 17 eV (magenta line), and the combined experimental and theoretical contribution which fills in the missing DR resonance strength in the energy range 17 -- 22 eV due to field ionization of high-$n$ Rydberg levels (dashed blue line).}
    \label{fig:prrc}
\end{figure*}

Our experimentally derived temperature-dependent plasma recombination rate coefficient $\alpha_\mathrm{exp}(T)$ is given in \autoref{fig:prrc}.

The total DR plasma recombination rate coefficient was constructed from the sum of three distinct components, namely: the low-energy fitted $\hat{\sigma}(E_\mathrm{cm})$, the overview composite spectrum in the range of 0.25 to 25 eV, and a correction for the missing DR resonance strength due to the field ionization of high-$n$ Rydberg levels in the storage ring dipole magnet behind the electron cooler.

The first component was constructed using the model cross section $\hat{\sigma}(E_\mathrm{cm})$ as given in \autoref{table:lowresfit} (also shown in \autoref{fig:ne3+fit}). Thereby, this quantifies how much recombination resonances below 0.25 eV contribute to the PRRC. This contribution is represented by the dotted green line in \autoref{fig:prrc}.

In the energy range 0.25 - 17 eV, the experimental DR rate coefficient was obtained by subtracting the theoretical radiative recombination contribution from the experimentally derived MBRRC as outlined in \citet{Lestinsky2009}. Subsequently, the resulting pure DR spectrum was convolved using \autoref{eqn:convolution} and its contribution to the PRRC is represented by the solid magenta line in \autoref{fig:prrc}.

As for the last component mentioned above, field ionization of recombined ions during their flight to the particle detector effectively truncates the measured DR spectrum above a maximum observable quantum number, $n_\mathrm{cut}$ = 23, leaving the high-$n$ Rydberg resonance contributions outside the measurable range. To account for this missing DR resonance strength, the experimental rate coefficient in the 22.0 -- 23.5 eV energy range was augmented by $ \alpha_{exp} + \kappa \cdot (\alpha^{\mathrm{AS}}_{1000} - \alpha^{\mathrm{AS}}_{n_{\mathrm{cut}}})$, following \citet{Schippers2004}, where $\kappa = 0.707$ (see \autoref{table:strengthdifferences}) is the scaling factor obtained by matching the \textsc{autostructure} calculation to the experimental spectrum, $\alpha^{\mathrm{AS}}_{1000}$ includes resonances up to $n = 1000$, and 
$\alpha^{\mathrm{AS}}_{n_{\mathrm{cut}}}$ is the calculation truncated at the 
field-ionization cut-off. The resulting extrapolation, representing the contribution 
of high-$n$ Rydberg resonances to the effective PRRC, is shown as the blue 
dash-dotted line in \autoref{fig:prrc}.

\begin{figure}[h]
    \hspace*{-0.35cm}
    \includegraphics[width=1.13\columnwidth]{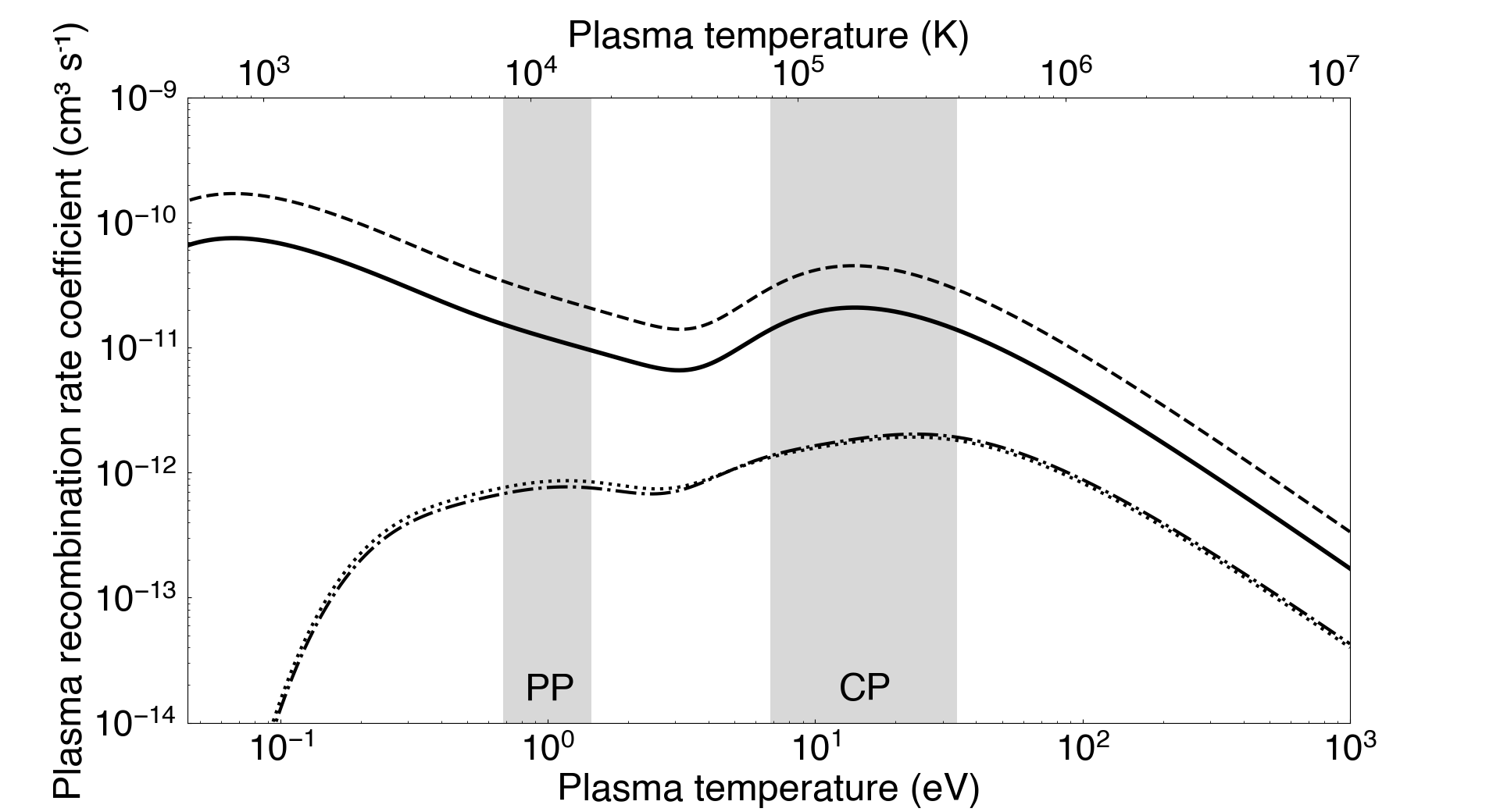}
    \caption{The theoretical data for each metastable state are listed separately by \citet{Badnell_webpage} and are presented as follows: the total PRRC, obtained by summing the contributions from the ground state and all metastable states according to their statistical weights, is represented by a thick solid line. The individual contributions are shown normalized to 100\% as thinner lines: the ground state $^4{S}_{3/2}$ (long-dashed), $^2{D}_{3/2}$ (dotted) and $^2{D}_{5/2}$ (short-dashed).}
    \label{fig:prrctheory}
\end{figure}

\autoref{fig:prrctheory} shows the theoretical PRRC \citep{Badnell_webpage} as a function of plasma temperature. The individual contributions from the ground state and each metastable state are weighted at 100\% and plotted separately. At all plasma temperatures, the total PRRC rate coefficient (thick solid line) represents the sum of the ground state contribution ($^4{S}_{3/2}$) (long-dashed line) and the two metastable state contributions ($^2{D}_{3/2}$: dotted and $^2{D}_{5/2}$: short-dashed) weighted accordingly. The relative weight of each contribution varies with temperature, reflecting the different energy dependencies of the underlying cross sections.

\begin{figure}[ht]
    \hspace*{-0.35cm}
    \includegraphics[width=1.13\columnwidth]{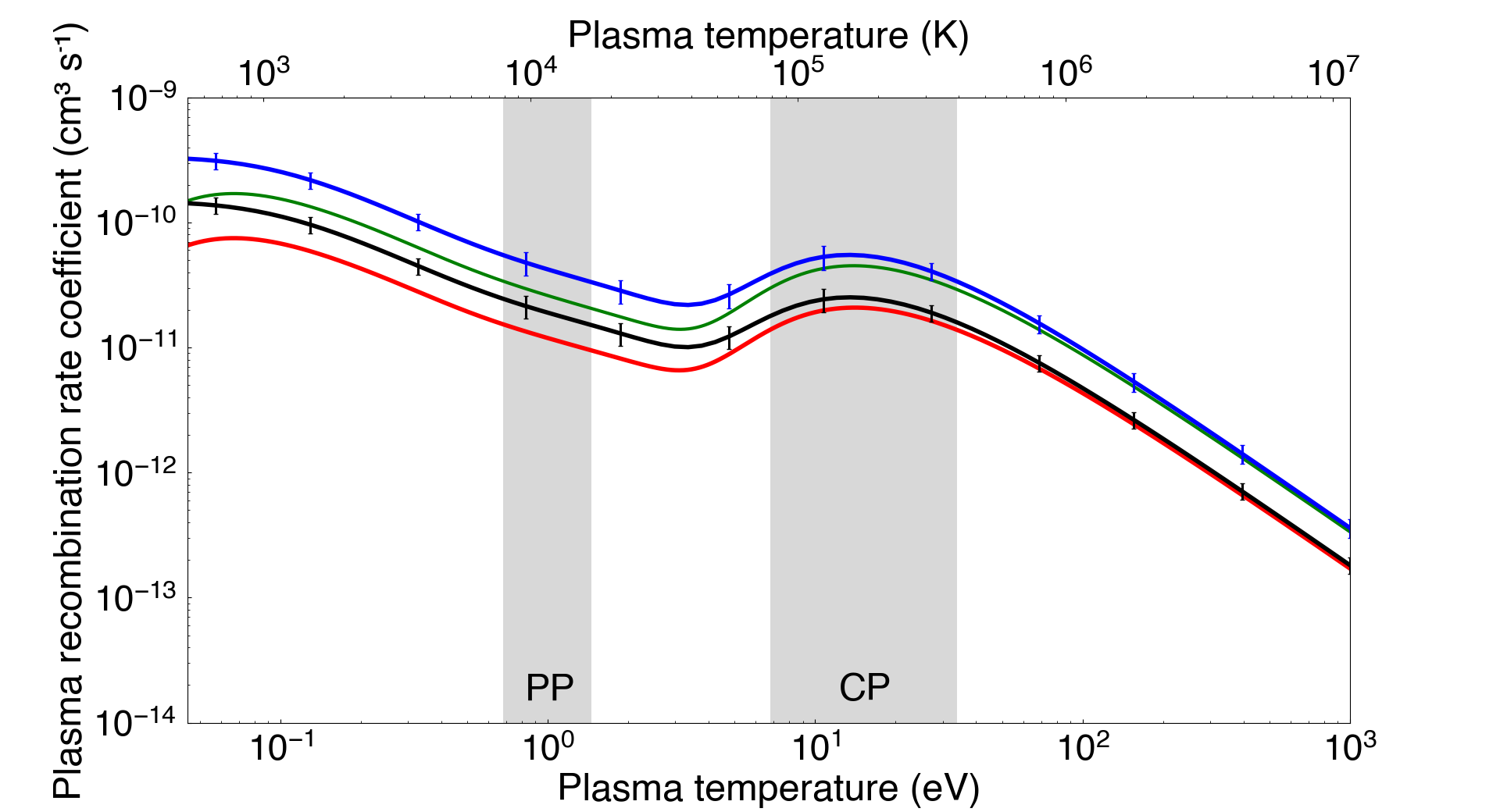}
    \caption{Comparison of our experimentally derived total PRRC (black line with error bars comprising statistical and systematic uncertainties) with the theoretical calculation of \citet{Badnell_webpage} that includes the weighted contributions of the metastable levels (red line). We also show a comparison between the experimentally derived PRRC considering only the ground-state contribution (blue line with associated error bars) and the theoretical calculation that likewise includes only the ground-state contribution (green line).}
    \label{fig:prrc_comp_theory}
\end{figure}

The sum of the three components described above (the total DR PRRC $\alpha_\mathrm{exp}(T)$), as well as the individual contributions, along with its associated total uncertainty, is derived and shown as a black thick solid line in \autoref{fig:prrc}. In \autoref{fig:prrc_comp_theory}, we provide a comparison of our experimental $\alpha_\mathrm{exp}(T)$ with theoretical results published on the ADAS website and provided by \citet{Badnell_webpage} via a simple parametrization. The red solid line corresponds to the weighted sum of ground and the individual metastable levels, while the green solid line considers only the contribution arising from the $^4{S}_{3/2}$ ground level.

The relative experimental uncertainties are 15.02\% at 0.01 eV, 20.81\% at 0.5 eV, 16.68\% at 17 eV and $\approx$15\% above 20 eV.

The experimentally derived ground-state-only PRRC (blue line in \autoref{fig:prrc_comp_theory}) is obtained by subtracting the metastable-level contributions from the total experimentally derived PRRC (black line). Removing these metastable contributions from the measured total PRRC therefore isolates the portion of the experimental PRRC attributable to the ground-state ion beam fraction alone, which can then be compared directly and consistently to the theoretical \citep{Badnell_webpage} ground-state-only calculation (green line). This comparison is complementary to the total PRRC comparison (black vs.\ red lines): while the latter tests the theory's ability to reproduce the full experimental beam composition, the former isolates the ground-state channel specifically as shown in \autoref{fig:prrc_comp_theory}. The theory-to-experiment discrepancy is nearly the same in both cases ($\sim$0.62 in the PP region and $\sim$0.77–0.78 in the CP region).

We parameterized the experimentally derived DR PRRC $\alpha_{T}$ by fitting the function:
\begin{equation}
    \alpha_{PRRC}^{fit} (T) = T^{-3/2} \sum_{i} c_{i} \exp(-E_i / T)
    \label{equation:prrcfit}
\end{equation}

The best-fit values of the free parameters $c_i$ and $E_i$ are listed in \autoref{tab:prrcfitcoef}. 
The same form was used to fit both the total experimentally derived PRRC and the experimentally derived ground-state-only PRRC (obtained as described above), the resulting fit coefficients for the total PRRC are given in the left-hand columns of \autoref{tab:prrcfitcoef}, and those for the ground-state-only PRRC in the right-hand columns.

\begin{table}[h!]
\caption{\label{tab:prrcfitcoef_combined}Fit parameters for the experimental DR PRRC of Ne$^{3+}$ using \autoref{equation:prrcfit}, for the total PRRC (represented by the black line in Figure~\ref{fig:prrc_comp_theory}) and for the PRRC assuming only ground-state contributions (represented by the blue line in the same figure).}
\centering
\begin{tabular}{lcccc}
\tableline
\multicolumn{1}{c}{} & \multicolumn{2}{c}{All levels} & \multicolumn{2}{c}{Ground level} \\
\tableline
\multicolumn{1}{c}{i} & $c_i$ & $E_i$ & $c_i$ & $E_i$ \\
\multicolumn{1}{c}{ } & \multicolumn{1}{c}{($\mathrm{cm}^3 \mathrm{s}^{-1} \mathrm{eV}^{3/2}$)} & \multicolumn{1}{c}{($\mathrm{eV}$)} & \multicolumn{1}{c}{($\mathrm{cm}^3 \mathrm{s}^{-1} \mathrm{eV}^{3/2}$)} & \multicolumn{1}{c}{($\mathrm{eV}$)} \\
\tableline
\multicolumn{1}{c}{1} & $8.938\cdot10^{-12}$ & $0.094$  & $1.752\cdot10^{-11}$ & $0.084$ \\
\multicolumn{1}{c}{2} & $7.3125\cdot10^{-12}$ & $0.497$ & $1.334\cdot10^{-11}$ & $0.336$ \\
\multicolumn{1}{c}{3} & $4.039\cdot10^{-11}$ & $1.902$  & $4.892\cdot10^{-11}$ & $1.394$ \\
\multicolumn{1}{c}{4} & $6.983\cdot10^{-11}$ & $5.165$  & $1.388\cdot10^{-10}$ & $3.459$ \\
\multicolumn{1}{c}{5} & $5.724\cdot10^{-9}$ & $21.63$   & $1.21\cdot10^{-8}$ & $21.113$ \\
\tableline
\tableline
\end{tabular}
\label{tab:prrcfitcoef}
\end{table}

The fit is applicable over a temperature range of $10^2$ to $10^7$ K. In this temperature range, the relative residuals R = $(\alpha_\mathrm{exp} (T) - \alpha_\mathrm{PRRC}^{fit} (T) ) / \alpha_\mathrm{exp} (T) $ remain within $\pm$0.35\%.

\section{Summary and conclusions}

Dielectronic recombination of Ne$^{3+}$ was measured at the CRYRING@ESR storage ring over the collision energy range 0 - 25 eV, yielding the merged-beams recombination rate coefficient $\alpha_\mathrm{exp}(E_\mathrm{cm})$. The experimental spectrum reveals the complete set of $\Delta n$ = 0 DR resonances associated with 2s $\rightarrow$ 2p core excitations from both the $^4S_{3/2}$ ground level and the low-lying metastable levels ($^2D_{5/2}$ and $^2D_{3/2}$). Approximately 44$\%$ of the stored Ne$^{3+}$ ions resided in the ground level, with the remainder distributed among long-lived metastable levels, as modelled from the ion-source population. The measured MBRRC is compared with \textsc{autostructure} calculations for the same weighted mixture of initial levels.
Between $\approx$9 and 25 eV the agreement in resonance strengths is within the combined experimental and theoretical uncertainties. At lower energies (0.5 – 9 eV), however, the experimental spectrum shows substantially stronger resonance clusters than predicted, leading to integrated strengths that substantially exceed the theoretical values. The dense resonance structure near the zero energy limit, dominated by contributions from the ground state series, was additionally characterized by a multi-resonance fit.
From the composite experimental spectrum, augmented by scaled \textsc{autostructure} results to account for high-$n$ Rydberg states (lost to field ionization), we derived the plasma recombination rate coefficient $\alpha_\mathrm{exp}(T)$ for Ne$^{3+}$ DR. In the temperature range characteristic of collisionally ionized plasmas, where Ne$^{3+}$ is abundant according to \citet{Bryans2009}, the experimentally derived PRRC shows good agreement with the weighted theoretical rate coefficients of \citet{Badnell_webpage}. By contrast, in the low-temperature regime relevant to photoionized plasma, such as planetary nebulae (0.68 -- 1.46 eV), our experimental PRRC is appreciably larger than the published theoretical values, by a factor of $\sim$1.6. This discrepancy is attributed primarily to the discrepancies between experimental and theoretical resonance strengths in the 0.5 -- 9 eV energy range.
We provide a ten-parameter analytic fit to the experimentally derived DR plasma rate coefficient that accurately reproduces $\alpha_\mathrm{exp}(T)$ from $10^2$ to $10^7$ K and is suitable for direct implementation in astrophysical plasma modelling codes.
Correct modelling of metastable-state contributions remains essential, as these levels may be populated both in the stored ion beam and in astrophysical environments. While the present results do not suggest a dramatic revision of the neon charge-state balance, the moderate increase in the low temperature DR rate may have an effect on the ionization correction factors and elemental abundance determinations derived from detailed photoionization models of planetary nebulae and other astrophysical plasmas at low temperatures. Nevertheless, future higher-statistics measurements, combined with refined theoretical treatments that better capture low-energy configuration-interaction effects, will further reduce the remaining uncertainties in the atomic data setting a foundation of neon diagnostics in nebular environments.

\begin{acknowledgments}

The results presented here are based on the experiment G-22-00047, which was performed at the heavy-ion storage ring CRYRING@ESR at the GSI Helmholtzzentrum f\"ur Schwerionenforschung, Darmstadt (Germany) in the frame of FAIR Phase-0 research program of the SPARC collaboration. We are grateful to the staff of the GSI Department of Accelerator Operations and Development for proficient support in preparing and maintaining the stored ion beam throughout the experiment. The Giessen group acknowledges financial support from the German Federal Ministry for Research, Technology, and Space (BMFTR) via the Collaborative Research Center ErUM-FSP T05 - \lq\lq Aufbau von APPA bei FAIR\rq\rq\ (Grant No.\ 05P24RG2). MRF acknowledges support from the National Science Foundation - Astronomy \& Astrophysical Research, \lq\lq Collaborative Research: A joint theoretical and experimental approach to low-temperature dielectronic recombination data for photoionized astrophysical environments\rq\rq). MRF acknowledges support from NASA - Astrophysical Research \& Analysis (\lq\lq A joint theoretical and experimental approach to dielectronic recombination data needs\rq\rq). This work has been conducted within the SPARC collaboration at FAIR and GSI and is part of the FAIR Phase-0 program.

\end{acknowledgments}

\begin{contribution}

ML, EBM, MRF and SS have prepared the experiment proposal.
EBM, EOH and ML have prepared the experimental setup.
AB, PMH, MDL, SS, MRF, RS, SXW and MT conducted the data taking under the supervision of ML, EBM and EOH. 
GV, WG, ZA, CK and FH have prepared and maintained the ion source and ring operation.
The experimental data were analysed by EOH.
SS and EOH provided the theoretical models.
This manuscript has been written by EOH and edited by all authors.

\end{contribution}

\bibliography{paper}{}
\bibliographystyle{aasjournalv7.bst}

\end{document}